# Thickness Measurements from Single X-ray Phase-contrast Speckle Projection

Yan Xi[1], Rongbiao Tang[2], Jingchen Ma[1], Jun Zhao[1,a)]

[1]School of Biomedical Engineering and Med-X Research Institute, Shanghai Jiao Tong University, Shanghai 200030, China

[2]Department of Radiology, Ruijin Hospital, School of Medicine, Shanghai Jiao Tong University, Shanghai 200030, China

**Abstract**

We propose a one-shot thickness measurement method for sponge-like structures using a propagation-based X-ray phase-contrast imaging (P-PCI) method. In P-PCI, the air-material interface refracts the incident X-ray. Refracted many times along their paths by such a structure, incident X-rays propagate randomly within a small divergent angle range, resulting in a speckle pattern in the captured image. We found structure thickness and contrast of a phase-contrast projection are directly related in images. This relationship can be described by a natural logarithm equation. Thus, from the one phase-contrast view, depth information can be retrieved from its contrast. Our preliminary biological experiments indicate promise in its application to measurements requiring *in vivo* and ongoing assessment of lung tumor progression.

**The manuscript**

X-ray imaging is a valuable tool in many areas as its investigation of objects is nondestructive. In X-ray images, interior structures appear overlapped and are represented by their X-ray absorption properties. To reconstruct a three-dimensional (3D) structure of the imaged object, multiple X-ray projections around the object are needed. However, this type of image processing cannot always be achieved because of concerns over radiation damage or procedures that cannot be repeated. To address this problem, we propose a one-shot thickness measurement method for sponge-like structures using X-ray phase-contrast imaging (PCI). With this method, depth information of an imaged sample can be retrieved from a single view. Thus, the 3D volume of a targeted object can be predicated.

X-ray PCI has been a hot topic for several years because of its fundamentally different imaging mechanism compared with traditional X-ray imaging[1, 2]. This technique is based on the phase shift associate with the incident X-rays passing through objects. A change in the X-ray wavefront results in a change in propagation direction. To represent the imaged object with phase-shift contrast, many imaging methods have been proposed, such as propagation-based PCI (P-PCI)[1], analyzer-based diffraction-enhanced imaging[3], and grating-based PCI (G-PCI)[2,4]. Among these, the P-PCI method offers the simplest system configuration and can achieve high spatial resolution as no optical elements are used within the X-ray path.

According to previous studies[5–8], air-material structures can introduce significant phase contrast for P-PCI. The sudden change in density at air-material interfaces can alter the propagation direction of X-rays. For example, refracted by air-filled blood vessels, bright lines are formed in a recorded phase-contrast image[9]; refracted by air-filled microbubbles, bright spots are formed[6, 8]. Considering that there are many air-material structures existing within an X-ray path, the direction of X-ray propagation changes randomly and often; speckle patterns thus form in projection images[7]. Phase-contrast imaging of sponge-like structures plays an important role in biomedical applications, such as lung tumor detection[10]. Normal alveoli act as mirrors, refracting X-rays that form speckle patterns in projection images. In contrast, dense cancerous tissue within the lung results in fewer fluctuations in tumor projections. This phenomenon has been studied to improve the contrast between normal and cancerous lung tissue[10, 11]. Because there are obvious differences between air-filled structures and dense materials, we developed a volume evaluation method to distinguish between sponge-like structures of differing thicknesses. Both model and biological experiments were performed to verify its effectiveness.

Without loss of generality, consider the two-dimensional interaction between X-rays and sponge-like structure during P-PCI. The imaging system is shown in Fig. 1(a), where the sample changes the propagation directions of incident planar X-rays. The refracted X-rays are captured by a detector downstream at a distance D away from the sample. During the interaction, each subunit, defined by a ball in our example, changes the X-ray propagation direction slightly, as indicated in Fig. 1(b), where $\theta_i$ denotes the angle of incident for the X-ray, and $\theta_{i+1}$ denotes its exiting direction. Exiting angles can be calculated using Snell's law. As this angle is only dependent on the incident angle and the interaction position with the subunit, but not its history propagation path, the interactions between X-rays and the many subunits in a sponge-like system can be analyzed as a Markov process. Let the difference in angle $\Delta\theta = \theta_{i+1} - \theta_i$ be described by probability distribution function $f$. As the ball is symmetrical, the probability distribution function of $\Delta\theta$ satisfies $f(\Delta\theta) = f(-\Delta\theta)$. Let $P^{(n)}_{\theta_i,\theta_j}$ be the probability function of a propagating X-ray changing direction from $\theta_i$ to $\theta_j$ in passing $n$ subunits, where $\theta \in \Omega$; here $\Omega$ is the X-ray direction space. According to the property of Markov chains, we have

$$
\begin{aligned}
P^{(n)}_{\theta_i,\theta_j} &= \sum_{r1 \in \Omega} P^{(n-1)}_{\theta_i,\theta_{r1}} P^{(1)}_{\theta_{r1},\theta_j} \\
&= \sum_{r1 \in \Omega} \left( \sum_{r2 \in \Omega} P^{(n-2)}_{\theta_i,\theta_{r2}} P^{(1)}_{\theta_{r2},\theta_{r1}} \right) P^{(1)}_{\theta_{r1},\theta_j} \\
&\cdots \\
&= \sum_{r1 \in \Omega} \sum_{r2 \in \Omega} \cdots \sum_{r(n-1) \in \Omega} P^{(1)}_{\theta_i,\theta_{r(n-1)}} P^{(1)}_{\theta_{r(n-1)},\theta_{r(n-2)}} \cdots P^{(1)}_{\theta_{r1},\theta_j} \\
&= \sum_{r1 \in \Omega} \sum_{r2 \in \Omega} \cdots \sum_{r(n-1) \in \Omega} f(\theta_i - \theta_{r(n-1)}) f(\theta_{r(n-1)} - \theta_{r(n-2)}) \cdots f(\theta_{r1} - \theta_j)
\end{aligned} \quad (1).
$$

As the incident X-rays are assumed plane waves $\theta_i = 0$, as depicted in Fig. 1(a), Eq. 1 can be written as

$$
\begin{aligned}
P^{(n)}_{0,\theta_j} &= \sum_{r1 \in \Omega} \sum_{r2 \in \Omega} \cdots \sum_{r(n-1) \in \Omega} f(-\theta_{r(n-1)}) f(\theta_{r(n-1)} - \theta_{r(n-2)}) \cdots f(\theta_{r1} - \theta_j) \\
&= \sum_{r1 \in \Omega} \cdots \sum_{r(n-2) \in \Omega} \left( \left( \sum_{r(n-1) \in \Omega} f(\theta_{r(n-1)}) f(\theta_{r(n-2)} - \theta_{r(n-1)}) \right) f(\theta_{r(n-2)} - \theta_{r(n-3)}) \cdots f(\theta_{r1} - \theta_j) \right) \\
&= \sum_{r1 \in \Omega} \cdots \sum_{r(n-2) \in \Omega} \left( (f \otimes f)(\theta_{r(n-2)}) f(\theta_{r(n-2)} - \theta_{r(n-3)}) \cdots f(\theta_{r1} - \theta_j) \right) \\
&\cdots \\
&= \underbrace{f \otimes f \otimes \cdots \otimes f}_{n-1}(\theta_j)
\end{aligned} \quad (2).
$$

Thus, the angular distribution of the X-ray direction after interaction with $n$ subunits is the convolution of ($n$-1) angular probability distribution functions. The result implies that the thicker the sponge-like structure, the wider the angular distribution for exiting X-rays. This conclusion can also be extended to interactions between X-rays and three-dimensional sponge-like structures.

In our model experiment, we fabricated a sponge-like structure using polymethyl methacrylate (PMMA) granules with a mean diameter ~120 μm. These granules were packed randomly with a density of around 52% between two glass slides and irradiated with parallel X-rays beams (Fig. 1(a)). The 3D visualization of the lab-made sponge-like structure is shown in Fig. 1(c). Imaging experiments were performed at the BL13W beamline of the Shanghai Synchrotron Radiation Facility with a partially coherent 25 keV X-ray beam. The distance between sample and detector, $D$, was set at 60 cm. Phase-contrast images were recorded using a thin (100 μm) $CdWO_4$ cleaved single-crystal scintillator and CCD camera with pixel size of 9 μm. One phase-contrast speckle projection of a 2 mm thick model is shown in Fig. 1(d).

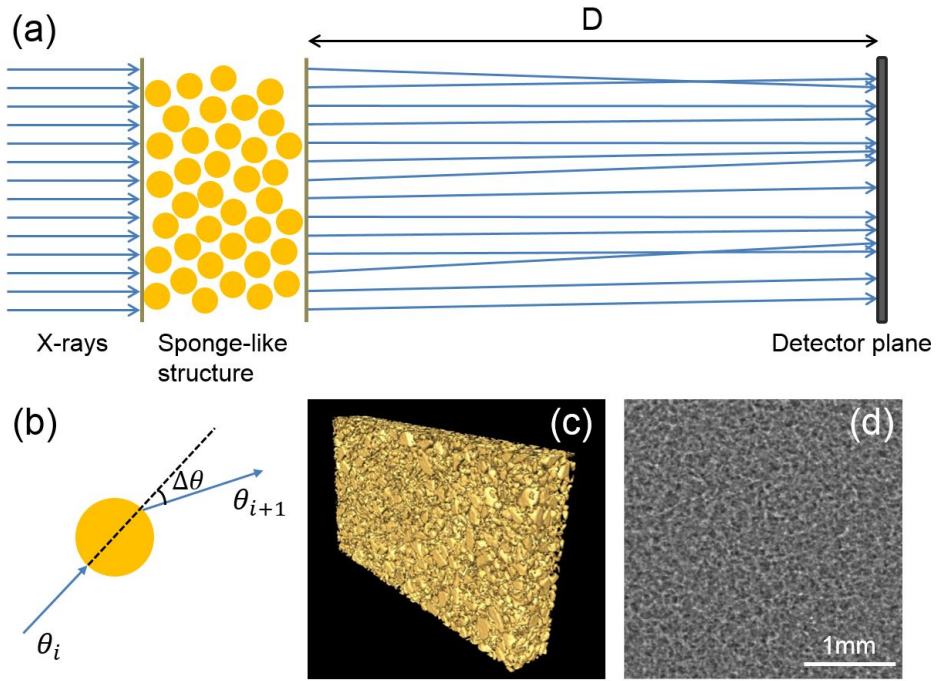

FIG. 1. (Color online) Illustrations of (a) propagation-based X-ray phase-contrast imaging and (b) interaction between a subunit and X-ray. (c) is the 3D visualization of the imaged lab-made sponge-like structure, and (d) its X-ray phase-contrast speckle projection.

To quantify the contrast, we employed

$$C(x,y) = \frac{\sqrt{\langle I(x,y)^2 \rangle_W - \langle I(x,y) \rangle_W^2}}{\langle I(x,y) \rangle_W} \qquad (3),$$

to calculate the contrast projection, which is composed of overlapping local contrasts[12]. In this formula, $(x, y)$ represents a point in the image, $I$ the intensity, subscript $W$ denotes the size of the local calculation window, and the operator $\langle \cdot \rangle$ signifies averaging over the local window. In our experiments, we adjusted the distance between the two glass slides to produce sponge-like structures of different thicknesses. Each granular model was randomly packed to maintain the same packing density; their P-PCI projections are listed in the left column in Fig. 2. Intuitively, these projections are darker for thicker sponge-like structures because of the absorption by materials. Furthermore, sharper fluctuations were observed in the P-PCI images with thicker structures. Using Eq. 3, quantified contrast projections were calculated in false color; typical images are placed alongside their P-PCI projections in the far-right column of Fig. 2. The size of the local calculation window is set at 30 pixels.

From our results for the different sample thicknesses, there are obvious color differences among the contrast images. By calculating, for each projection, the mean speckle contrast and its standard deviation, and plotting against sample thickness (Fig. 3(a)), a relation can be established modeled with a natural logarithm equation $C = a \times \ln(b \times T + 1)$, where $a$ and $b$ are parameters. A fit to the data yields $a = 0.0378$ and $b = 4.7148$, with an adjusted R-square of 0.9957.

In Eq. 1, the calculation window size affects the spatial resolution of the quantified contrast image; the larger the calculation window, the lower the spatial resolution. To study the robustness of speckle contrast with different calculation window sizes, we plotted the speckle contrast of projection of 1.5 mm sponge-like structures with various window sizes ranging from 10 to 50 pixels (Fig. 3(b)). The result suggests that the quantified contrast maintains stability with different calculation windows. However, with increasing calculation window size, the standard deviation for the speckle contrast decreases, indicating a flatter contrast image. There is a trade-off between spatial resolution and noise in the quantified contrast image.

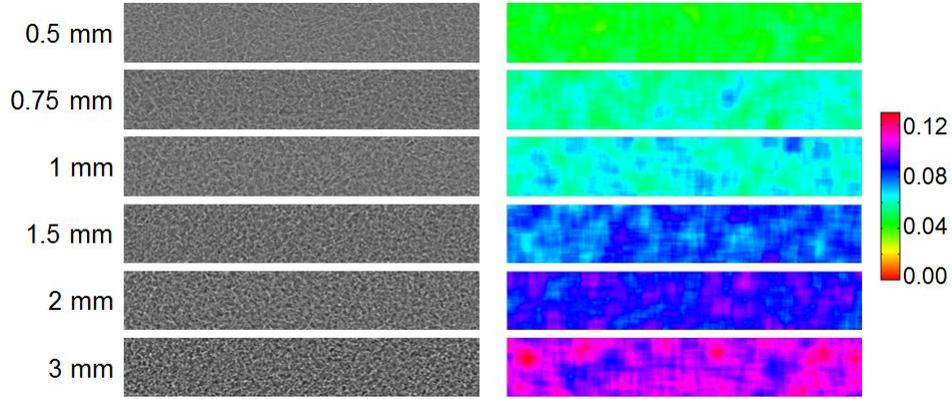

FIG. 2. (Color online) (left column) X-ray phase contrast projections for samples of different thicknesses and (right column) their corresponding quantified contrast images.

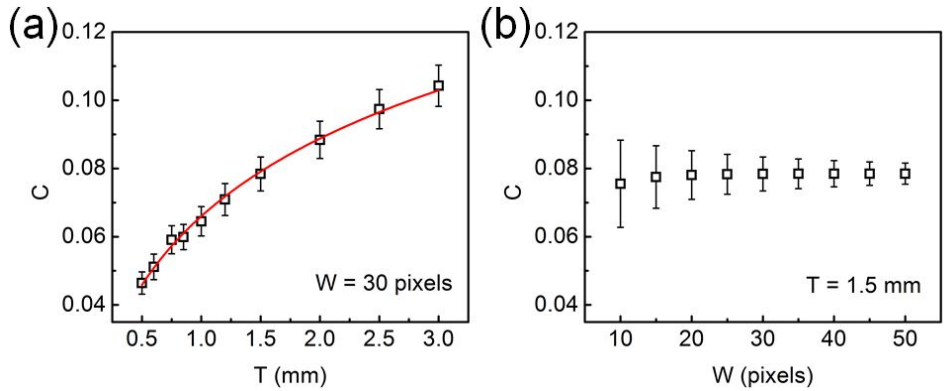

FIG. 3. (Color online) (a) Relationship between speckle contrast and sample thickness with a fixed calculation window of 30 pixels. (b) Relationship between speckle contrast and calculation window size with a 1.5 mm sample thickness. Each data point is expressed as mean ± standard deviation in the corresponding phase-contrast projection. The red line in (a) is the result of fitting with the natural logarithm equation.

In our biological experiment, two male C57BL/6 mice (about 6 weeks old) were studied *in vivo*. Both were implanted with Lewis lung tumors, one with 3-day-old cells and the other one with 5-day-old cells. Their tumor thicknesses of about 1.68 and 3.40 mm. The imaging configuration is the same as in the model experiment described above except that the X-ray energy is 20 keV. P-PCI projection images of the left lung for each mouse are presented (Fig. 4(a) and (b)). The tumor areas are framed by white boxes. Enlarged views of these areas are presented in Fig. 4(f) and (g) with a narrow display window. Tumors are indicated by white arrows. A simple sketch of a tumor inside lung tissue (Fig. 4(e)) is presented as an aid to visualize the model, with hollow circles representing normal alveoli and the solid ellipse representing the tumor of thickness $T_2$; $T_1$ denotes the thickness of whole line tissue. According to the above results, the speckle contrast is related to the thickness of the sponge-like structure ($T_1 - T_2$). Thus, by taking the phase-contrast projection of a lung tumor area and calculating its speckle contrast, the 3D size of the tumor can be predicated from one view.

In a further experiment, the lung tissue used in imaging (Fig. 4(a) and (b)) was manually segmented by removing the ribs; their quantified speckle contrast is shown in Fig. 4(c) and (d). Because the contrast is affected by many factors, such as respiratory phase, we employed instead the normalized speckle contrast expressed as

$$C_N(x,y) = \frac{C(x,y)}{<C_{base}>},$$

where $C_{base}$ is the reference speckle contrast. For this purpose, we calculated the speckle contrast of the right lung tissue of each mouse. The mean ± standard deviation normalized speckle contrasts of lungs with 3- and 5-day-old tumors were 0.53 ± 0.10 and 0.38 ± 0.08, respectively.

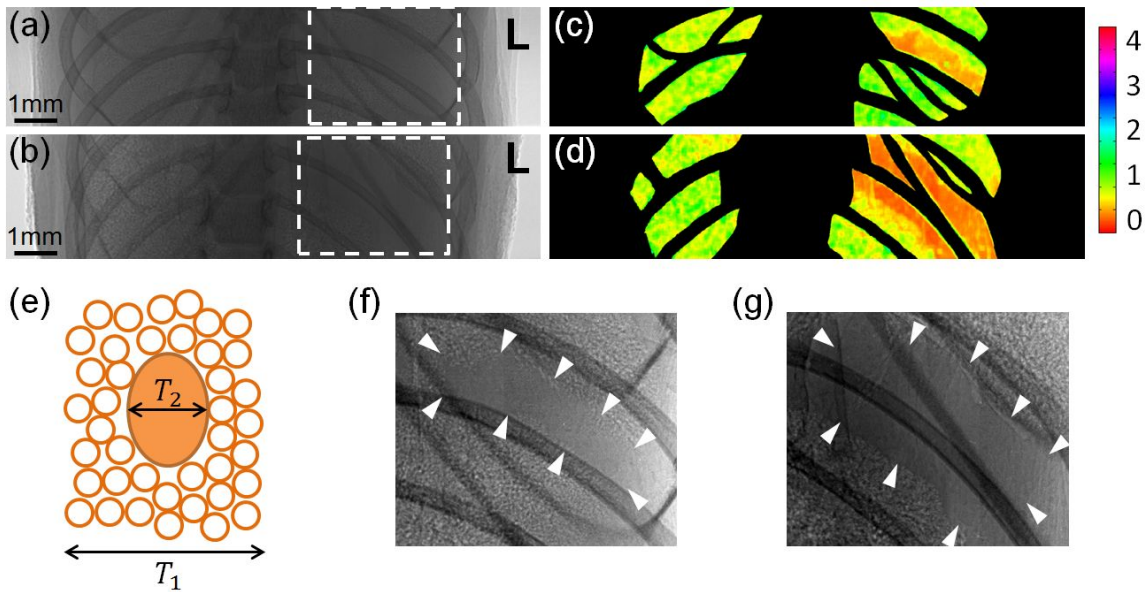

FIG. 4. (Color online) (a) and (b) are the respective *in vivo* phase-contrast projection images of the left lungs of the mice with 3- and 5-day-old tumors. ("L" signifies left side); (c) and (d) are their corresponding normalized speckle contrast images obtained using Eq. 2; (f) and (g) show the enlarged tumor areas box-framed in (a) and (b) with white arrows indicating tumors: (e) is a simple sketch of a lung tumor surrounded by alveoli cells where $T_1$ and $T_2$ denote the thickness of lung tissue and dense tumor tissue, respectively.

As demonstrated by our model experiment, the contrast in a phase-contrast projection from sponge-like materials is directly related to its thickness (Fig. 3(a)). From the contrast-thickness curve, we can determine the thickness of an imaged sample from its speckle contrast. The results are an interesting discovery that enables 3D measurements from a single projection image. Previous studies have shown improved image contrasts between normal lung tissues and lung tumors[10]. In our preliminary animal experiment, the quantified lung tumor projections corresponding to different development stages are represented in false color scale (Fig. 4(c) and (d)). This result suggests that the proposed one-shot thickness measurement can be potentially used *in vivo* to assess the progression of lung tumors. Because only one projection is needed and because of the high flux of synchrotron radiation, this method benefits from low radiation dosage and fewer artifacts caused through breathing. To have a systematic study on the relationship between lung tumor thickness and normalized speckle contrast, more animal experiments need to be performed.

In conclusion, we described a volume measurement method that only requires a single projection image. In our model experiment, we found that the contrast of the phase-contrast projection was directly related to sample thickness and expressible by a natural logarithm equation. Our preliminary biological experiment has shown that the one-shot volume measurement method can potentially be used for volume measurements of lung tumors, particularly taken *in vivo* for continuous monitoring of lung tumor growth. Further work is in progress.

**Acknowledgements**

This work was performed at the BL13W of the Shanghai Synchrotron Radiation Facility and supported by the National Basic Research Program of China (973 Program; 2010CB834302).

a)junzhao@sjtu.edu.cn

**References**

1. S. Wilkins, T. Gureyev, D. Gao, A. Pogany, and A. Stevenson, Nature **384** (6607), 335 (1996).
2. F. Pfeiffer, T. Weitkamp, O. Bunk and C. David, Nat. Phys. **2** (4), 258-261 (2006).
3. D. Chapman, W. Thomlinson, R. Johnston, D. Washburn, E. Pisano, N. Gmür, Z. Zhong, R. Menk, F. Arfelli, and D. Sayers, Phys. Med. Biol. **42** (11), 2015 (1999).
4. A. Momose, Opt. Express **11** (19), 2303 (2003).
5. R. Lewis, N. Yagi, M. Kitchen, M. Morgan, D. Paganin, K. Siu, K. Pavlov, I. Williams, K. Uesugi, and M. Wallace, Phys. Med. Biol. **50** (21), 5031 (2005).
6. Y. Xi, R. Tang, Y. Wang, and J. Zhao, Appl. Phys. Lett. **99** (1), 011101 (2011).
7. M. Kitchen, D. Paganin, R. Lewis, N. Yagi, K. Uesugi, and S. Mudie, Phys. Med. Biol. **49** (18), 4335 (2004).
8. R. Tang, Y. Xi, W.-M. Chai, Y. Wang, Y. Guan, G.-Y. Yang, H. Xie, and K.-M. Chen, Phys. Med. Biol. **56** (12), 3503 (2011).
9. Y. Xi, X. Lin, F. Yuan, G.-Y. Yang and J. Zhao, Internal Report (2013).
10. P. Liu, J. Sun, Y. Guan, W. Yue, L. X. Xu, Y. Li, G. Zhang, Y. Hwu, J. H. Je, and G. Margaritondo, J. Synchrotron Radiat. **15** (1), 36 (2007).
11. X. Liu, J. Zhao, J. Sun, X. Gu, T. Xiao, P. Liu, and L. X. Xu, Phys. Med. Biol. **55** (8), 2399 (2010).
12. M. Draijer, E. Hondebrink, T. Van Leeuwen, and W. Steenbergen, Lasers Med. Sci. **24** (4), 639 (2009).